\documentclass{article}
\usepackage{spconf,amsmath,graphicx}
\usepackage{xcolor}
\title{Image Processing in DNA}
\name{\begin{tabular}{c}Chao Pan$^{1}$, S. M. Hossein Tabatabaei Yazdi$^{2}$, S Kasra Tabatabaei$^{1}$\\ Alvaro G. Hernandez$^{1}$,
Charles Schroeder$^{1}$, Olgica Milenkovic$^{1}$\sthanks{The work was funded by the DARPA Molecular Informatics Program, The NSF+SRC SemiSynBio program under agreement number 1807526 and the NSF grant 1618366.} \end{tabular}}
\address{$^{1}$University of Illinois, Urbana-Champaign, $^{2}$Dorna Robotics \\
}
\vspace{-0.1in}
\begin{document}
%\ninept
%
\maketitle
\begin{abstract}
%DNA-based data storage is an emerging nonvolatile storage technology which offers unprecedented information densities and media lifetimes.
The main obstacles for the practical deployment of DNA-based data storage platforms are the prohibitively high cost of synthetic DNA and the large number of errors introduced during synthesis. In particular, synthetic DNA products contain both individual oligo (fragment) symbol errors as well as missing DNA oligo errors, with rates that exceed those of modern storage systems by orders of magnitude. These errors can be corrected either through the use of a large number of redundant oligos or through cycles of writing, reading, and rewriting of information that eliminate the errors. Both approaches add to the overall storage cost and are hence undesirable. Here we propose the first method for storing quantized images in DNA that uses signal processing and machine learning techniques to deal with error and cost issues without resorting to the use of redundant oligos or rewriting. Our methods rely on decoupling the RGB channels of images, performing specialized quantization and compression on the individual color channels, and using new discoloration detection and image
inpainting techniques. We demonstrate the performance of our approach experimentally on a collection of movie posters stored in DNA.
\end{abstract}
\section{Introduction}
\label{sec:intro}

DNA-based data storage has recently emerged as a viable alternative to classical storage devices that can be used to record bits at a nanoscale level and preserve them in a nonvolatile fashion for thousands of years~\cite{church2012next,goldman2013towards,yazdi2015rewritable,grass2015robust,yazdi2017portable,zhirnov2016nucleic,erlich2017dna,yazdi2015dna,tabatabaei2020dna}. Almost all existing DNA-based data recording architectures store user content in synthetic DNA strands of length $100-1000$ basepairs, organized within large unordered pools, and retrieve desired information via next-generation (e.g., HiSeq and MiSeq) or third-generation nanopore sequencing. Although DNA sequencing can be performed at a very low cost, de novo synthesis of DNA oligos with a predetermined content still represents a major bottleneck of the platform. Synthetic DNA platforms are prohibitively expensive compared to existing optical and magnetic media~\cite{zhirnov2016nucleic}. Furthermore, synthetic DNA-based storage systems have error-rates of the order of $10^{-3}$ that by far exceed those of existing high-density recorders. Synthesis errors include both symbol errors as well as \emph{missing oligo errors} which are unique to this type of storage media and refer to the fact that one may not be able to cover all substrings of the user-defined string. Missing oligos represent serious obstacles to accurate data retrieval as they may affect more than $20\%$ of the product. To address this type of error, Grass et al.~\cite{grass2015robust} proposed using Reed-Solomon codes at both the oligo and pool of oligo level to ensure that missing strings may be reconstructed from combinations of redundantly encoded oligos. Unfortunately, adding redundant oligos further increases the cost of the system as the oligos have to be sequenced to determine the missing oligo rate in order to add the correct amount of redundancy.
%\vspace{-0.1in}
%This sequencing process further adds to the cost of the storage device implementation. %\textcolor{blue}{Take twist result as an example?}.
%\begin{figure}[htb]
%\centerline{\includegraphics[width=8.5cm]{missing_oligos.PNG}}
%  \vspace{2.0cm}
%\caption{Example of the missing oligo phenomena: Parts of the user-defined sequence may not be present in the pool.}
%\label{fig:missing_oligo}
%\end{figure}
%\vspace{-0.1in}

We propose a new means of archiving images in DNA in which the missing and erroneous oligos are corrected through specialized learning methods, rather than expensive coding redundancy. The gist of our approach is to first aggressively quantize and compress colored images by specialized encoding methods that separately operate on the three color channels. Our quantization scheme reduces the image color pallet to $8$ intensity levels per channel, and compresses intensity levels through a combination of Hilbert-space filling curves, differential and Huffman coding. Given that compression may lead to catastrophic error-propagation in the presence of missing or mismatched oligos, we also introduce very sparsely spaced markers into the oligo codes in order to resynchronize positional pixel information when this is lost. No error-correcting redundancy is added to the pool in order to further save in synthesis cost, and instead, the retrieved corrupted images are subjected to specialized image processing techniques that lead to barely distorted outputs. Our scheme combines automatic detection of discolorations in images with inpainting based on EdgeConnect~\cite{Nazeri_2019_ICCV} and smoothing via bilateral filtering~\cite{tomasi1998bilateral}. It is worth pointing out that our scheme has no commonalities with joint source-channel coding approaches for erasure channels~\cite{balakirsky1997joint} as it explicitly avoids the channel coding component. We experimentally tested our proposed DNA image processing scheme on a pool of $11,826$ oligos of length $196$ basepairs each.
%The oPools contain eight Marlon Brando movie posters, one of which is used as the running example and has $10/1981$ missing oligos.
% we use as our running example. The initial pool of oligos capturing the image missed $10/1981$ oligos, respectively. %and were processed using the above described scheme to arrive at images with barely visible discolorations.

The paper is organized as follows. Section~\ref{sec:enc} contains a description of our color image encoding scheme, while Section~\ref{sec:amp} describes the oligo synthesis, amplification, image processing procedures and experimental results.

\vspace{-0.1in}
\section{The encoding procedure}
\label{sec:enc}

Our two-step encoding procedure first translates an image file into $24$ binary strings, and then converts the binary strings into DNA oligos for storage and amplification. A detailed description of each step used in the process is provided below.

\textbf{Converting image files to 2D arrays.} %Before encoding, we make an assumption that local smoothness exists in images, meaning that for each pixel it should share similar values with some of other pixels in neighborhood. In order to make full advantage of local smoothness property to reduce redundant information, five sub-steps are specially designed to convert a photo into 24 binary strings.
The first step in the procedure is \emph{RGB channel separation and quantization.} Some previous work~\cite{dimopoulou2019biologically} proposed to store quantized wavelet coefficients information, while we are directly quantizing pixel intensities. First, we split the color images into three color channels, R (red) G (green) B (blue), and then perform $3$-bit quantization of the values in each channel. More precisely, the image $I$ is represented by a three-dimension tensor of size $m\times n\times 3$, i.e., $I\in [256]^{m\times n\times 3}$, which we split  into three matrices $\textbf{R}, \textbf{G}, \textbf{B}$ of size $m\times n$ each. Next, we perform $3$-bit quantization of each color matrix, leading to intensity values mapped from $0-255$ to $0-7$. More specifically, we use the following quantization rule for all three channels:
\begin{equation}
\textbf{X}[p,q] = \text{floor}\left(\frac{\textbf{X}[p,q]\times 8}{256}\right)\quad \forall p\in[m],q\in[n],
\end{equation}
where $\textbf{X}\in[8]^{m\times n}$ is the quantized matrix for $\textbf{X} \in \{{\textbf{R},\textbf{G},\textbf{B}\}}$. 
%The same procedure can be done for green and blue channel, and we can get quantized matrices $R,G,B$.

\begin{figure}[htb]
\centerline{\includegraphics[width=5.1cm]{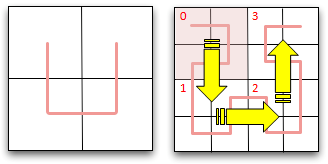}}
%  \vspace{2.0cm}
\caption{Hilbert curves for $2\times 2$ and $4\times 4$ squares.}
\label{fig:hilbert}
\end{figure}

\textbf{Converting 2D arrays to 1D binary strings.} There exist several methods for converting a matrix into a string so as to nearly-optimally preserve 2D image distances in the 1D domain, such as the Hilbert and Peano space-filling curve. The Hilbert space-filling curve, shown in Figure~\ref{fig:hilbert}, provides a good means to capture 2D locality~ \cite{hilbert1935stetige,moon2001analysis} and is the method of choice in our conversion process. Note that the Hilbert curve is standardly used on square images, so we adapt the transversal implementation to account for matrices with arbitrary dimensions (the detailed description of the mapping is postponed to the full version of the paper). After the mapping, the matrices $\textbf{R},\textbf{G},\textbf{B}$ are converted into vectors $V_R,V_G,V_B$, respectively.
%For example, a $2\times 2$ matrix mapping would be
%$$
%\textbf{R} = \begin{bmatrix}
%1 & 2\\
%3 & 4
%\end{bmatrix}\longrightarrow V_R = [1, 3, 4, 2].
%$$
% \vspace{-0.1in}

\textbf{Partitioning according to intensity levels.} Upon quantization, the values in $V_R,V_G,V_B$ lie in $\{{0,\ldots,7\}}$. We next decompose each vector into strings of possibly different lengths according to the intensity value. Specifically, $V_R$ is decomposed into $L_{R,0},\dots,L_{R,7}$, where the vector $L_{R,j}$ contains the indices of the elements in $V_R$ whose value equals $j$, $j\in[8]$; the same procedure is performed for the vectors $V_G,V_B$. An example decomposition may read as:
$$
V_R=[0,0,0,1,7,6,7,7,\dots] \longrightarrow
\begin{matrix}
L_{R,0} = [0,1,2,\dots] \\
\vdots \\
L_{R,7} = [4,6,7,\dots]
\end{matrix}
$$
Note that the elements in $V_i$ are assigned to $L_{i,j}$ in order, $i\in\{R,G,B\}$, $0\leq j\leq 7$. Hence each vector $L_{i,j}$ contains increasing values, a fact that we exploit in our reconstruction procedure. Given the Hilbert scan, one may expect the differences between adjacent entries in each of the vectors $L_{i,j}$ to be small with high probability. Therefore, splitting a vector into individual levels enables subsequent differential encoding~\cite{porter2007differential}. Moreover, since the level information is split among different vectors, we will also be able to correct distortions in the images in the presence of errors. In summary, after the RGB decomposition and level partition, each image is represented by $24$ vectors. \textbf{Differential encoding} converts a string into another string containing the initial value of the original and the differences between consecutive values, summarized in vectors denoted by $D_{i,j}$. In order to prevent catastrophic error propagation, we set $3\%$ of the values in each $D_{i,j}$ to their original undifferentiated values and prepend to them the symbol $-1$. We also append an additional $-2$ to each $D_{i,j}$ to indicate the end of the vector. For example, a typical pair of $L_{i,j}$ and $D_{i,j}$ may be of the form:
$$
\begin{matrix}
L_{i,j} = [x_1,x_2,\dots,x_{31},x_{32},\dots] \\
\downarrow \\
D_{i,j} = [-1,x_1,x_2-x_1,\dots,-1,x_{31},x_{32}-x_{31},\dots,-2].
\end{matrix}
$$
Note that as $L_{i,j}$ has increasing values, the symbols $-1$ and $-2$ cannot be confused with information-bearing values in $D_{i,j}$. %In decoding procedure $-1$s work as tags to indicate groundtruth positions, so algorithm can update correct accumulated value in case error happens.
\textbf{Huffman coding}~\cite{huffman1952method} is performed after differential coding, and all values in $D_{i,j}$ are used to construct the Huffman code dictionary. This results in a collection of binary strings $B_{i,j},$ $i\in\{R,G,B\}$, $0\leq j\leq 7$.

\textbf{Converting binary strings to DNA oligos.} The binary information is then converted into oligo strings over the alphabet $\{{\text{A,T,G,C}\}}$. To ensure a high quality of the synthetic product, we perform constraint coding by imposing a maximum runlength-$3$ constraint for the symbols C and G and ensuring a GC content in the range $40-60\%$~\cite{yazdi2015rewritable}. The constrained coder maps $18$ and $22$ bits binary strings into $10$ and $13$ nts DNA blocks, respectively. Each designed DNA oligo is of length 196 nts and is parsed into the following subsequences: A pair of primer sequences, each of length 20 nts, used for the prefix and suffix, an address sequence of length 10 nts, and 11 information-bearing sequences of length 13 nts.

\textbf{Primer sequences.} We add to each DNA oligo a prefix and suffix primer, used for PCR amplification of the single stranded DNA oligos. To allow for random access, we choose $8$ pairs of primers of length $20$, one for each level, all of which are at a Hamming distance $\geq 10$ nucleotides. The primers are paired up so as to have similar melting temperature, which allows for all oligos to be amplified in the same cycle.

\textbf{Address sequences.} Strings of length $13$ are added to the DNA oligos following the primers in order to represent the address of the information blocks contained. The first $3$ nucleotides of the address encode the color (RGB).
Since color information is highly important for reconstruction, we present it in redundant form as R = 'ATC', G = 'TCG', B = 'GAT'. This allows for single-error correction in the color code. The second part of the address is of length $10$ nucleotides, encoding a $18$-bit binary string including the index of the corresponding image file, the index of the color level and the index of the information block within that level.
%With address sequences, we can easily assemble information stored in different oligos.

\begin{figure}[htb]
\centerline{\includegraphics[width=8.5cm]{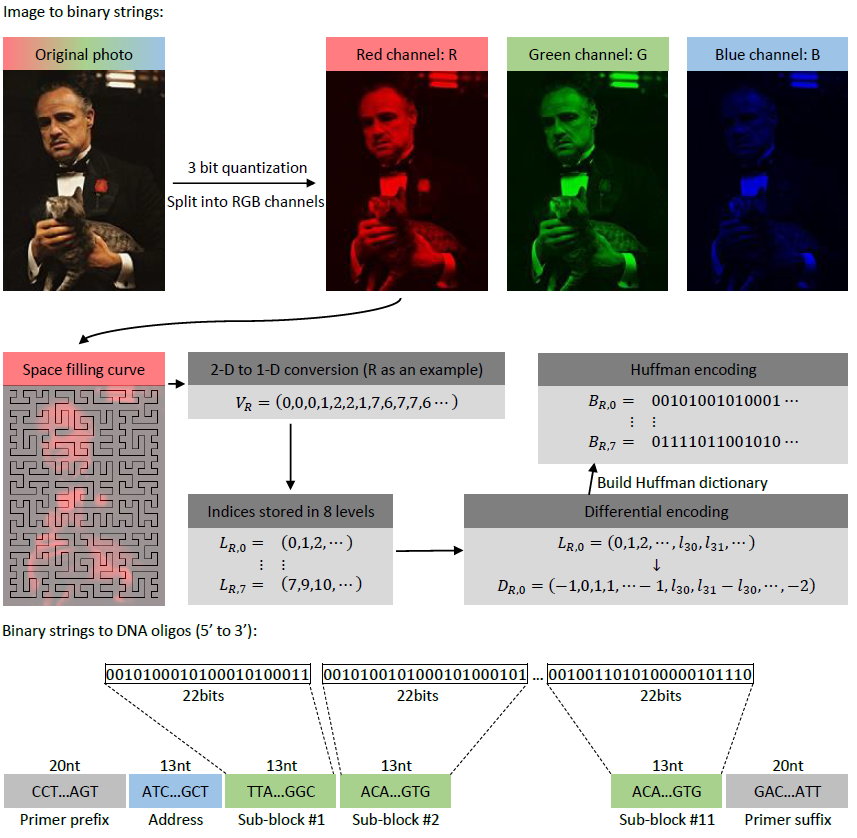}}
\caption{Schematic depiction of the encoding procedure.}
\label{fig:enc}
\end{figure}

\textbf{Information blocks} are added to the oligos between the address and suffix primer, including $11$ blocks of length $13$ nucleotides. The total length of the information block is $143$ nucelotides. Overall, with the compression scheme and additional addressing information added, $8,654,400$ bits of the original images are converted into $2,317,896$ nucleotides. The encoding steps are summarized in Figure~\ref{fig:enc}.
%And the comparison of storage space is shown in Table~\ref{tab:storage}.
%\begin{table}
%\begin{center}
%\begin{tabular}{ |c|c|c| }
% \hline
% &Original color image & DNA oligos\\
% \hline
% \tabincell{c}
%  {Space needed \\
%  for storage} & $8,654,400$ bits & $2,317,896$ nt\\
% \hline
%\end{tabular}
%\end{center}
%\vspace{-0.1cm}
%\caption{Storage space comparison.}
%\label{tab:storage}
%\end{table}
\vspace{-0.1in}

\section{DNA image processing and experiments}
\label{sec:amp}
The $11,826$ DNA oPools oligos were ordered from IDT (https://www.idtdna.com/pages/products/custom-dna-rna/dna-oligos/custom-dna-oligos/opools-oligo-pools). They were PCR amplified and the PCR products were then converted into a shotgun sequencing library with the Hyper Library construction kit from Kapa Biosystems (Roche). The library was quantitated by qPCR and sequenced on one ISeq flowcell for $251$ cycles from one end of the fragments. The fast file was generated with the Illumina bcl2fastq v2.20 Conversion Software. As each oligo read may contain errors that arise both during synthesis and sequencing, we first \emph{reconstructed a consensus sequence} via sequence alignment  to exploit the inherent redundancy of the read process. After the whole writing, reading and consensus process, we obtained $10,981$ perfectly reconstructed oligos, $745$ oligos with symbol errors that do not cause obvious defects in the reconstructed images, and $100$ oligos with large corruption levels or completely missing from the pool.
%The missing-oligo problem can be solved by repeating the writing and reading process, but results reported in this paper comes from the first round order with many errors.
%\section{Decoding procedure}
%\label{sec:dec}

%Binary strings are translated from DNA oligos, then images are decoded from those binary strings. Specifically,  first we decode binary user information from information blocks within DNA oligos, then we concatenate those strings according to their address sequences. Next for each long binary string, we need to do Huffman decoding, differential decoding, then assemble different level vectors having the same file id and color to one vector, map those vectors to 2D matrices, and finally merge RGB channels to obtain reconstructed color images.
%If no error occurs after consensus, meaning that all encoded oligos are perfectly matched after writing and reading process, we can obtain exact quantized images that we stored in DNA oligos. Furthermore, some decoding strategies introduced in following parts are used to handle possible errors after consensus.

\textbf{Converting DNA consensus sequences into binary strings.} The decoding procedure operates on the consensus reads and reverses the two-step encoding process. A detailed description of the procedure is omitted here. Due to synthesis and sequencing errors, we may not be able to find some identifiers in consensus reads for reversed constrained coder map during decoding. Thus we will replace erroneous identifiers by existing identifiers at smallest Hamming distance from it. In this case any DNA block can be converted into some binary string, although this string may be wrong and cause visible discolorations in the image.

\begin{figure}[htb]
\centerline{\includegraphics[width=7.5cm]{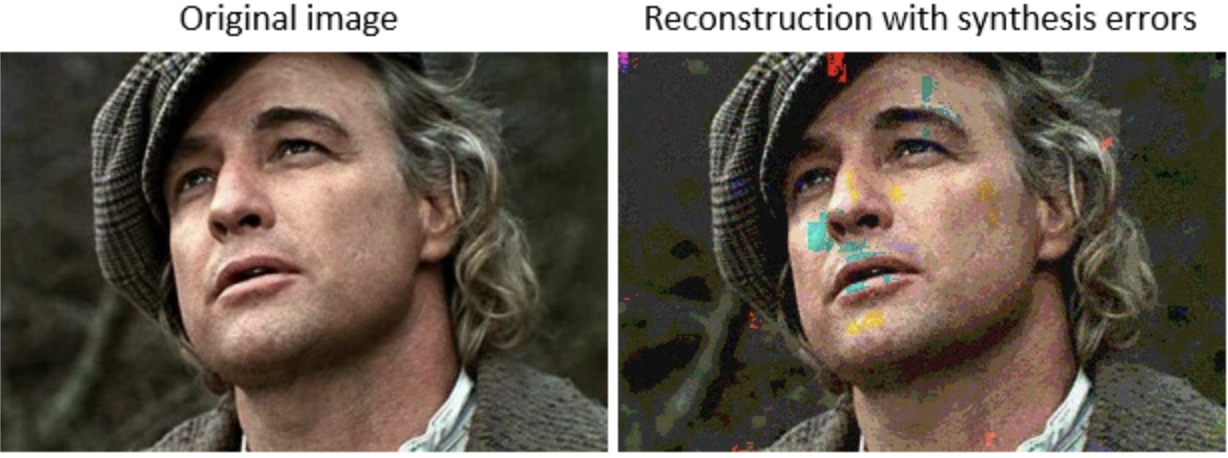}}
\caption{The reconstructed image with discolorations.}
\label{fig:ori}
\end{figure}

\textbf{Image processing.} An example illustrating the image corruptions caused by erroneous/missing oligos is shown in Figure~\ref{fig:ori}. Small blocks with the wrong color can be easily observed visually, and they are a consequence of only $10$ missing oligos. To correct the discolorations automatically, we propose a three-step image processing procedure. The first step consists of detecting the locations with discolorations, masking the regions with discolorations and subsequently treating them as missing pixels. The second step involves using deep learning techniques to inpaint the missing pixels. The third step involves smoothing the image to reduce both blocking effects caused by aggressive quantization and the mismatched inpainted pixels. %Detailed description of each step is introduced as follows.

%\textcolor{red}{Chao, since we have so little space, maybe it would be best to remove the second image with both Brando and Vivien altogether? I already changed the text to reflect that we only have one example image.}

\begin{figure}[htb]
\vspace{-0.1in}
\centerline{\includegraphics[width=6.9cm]{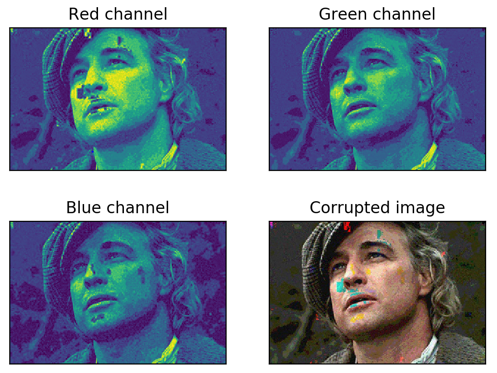}}
\vspace{-0.3cm}
\caption{Non-overlapping errors in different color channels of the image encoded in DNA.}
\label{fig:rgb_sep}
\vspace{-0.1in}
\end{figure}

\begin{figure}[htb]
\centerline{\includegraphics[width=5.8cm]{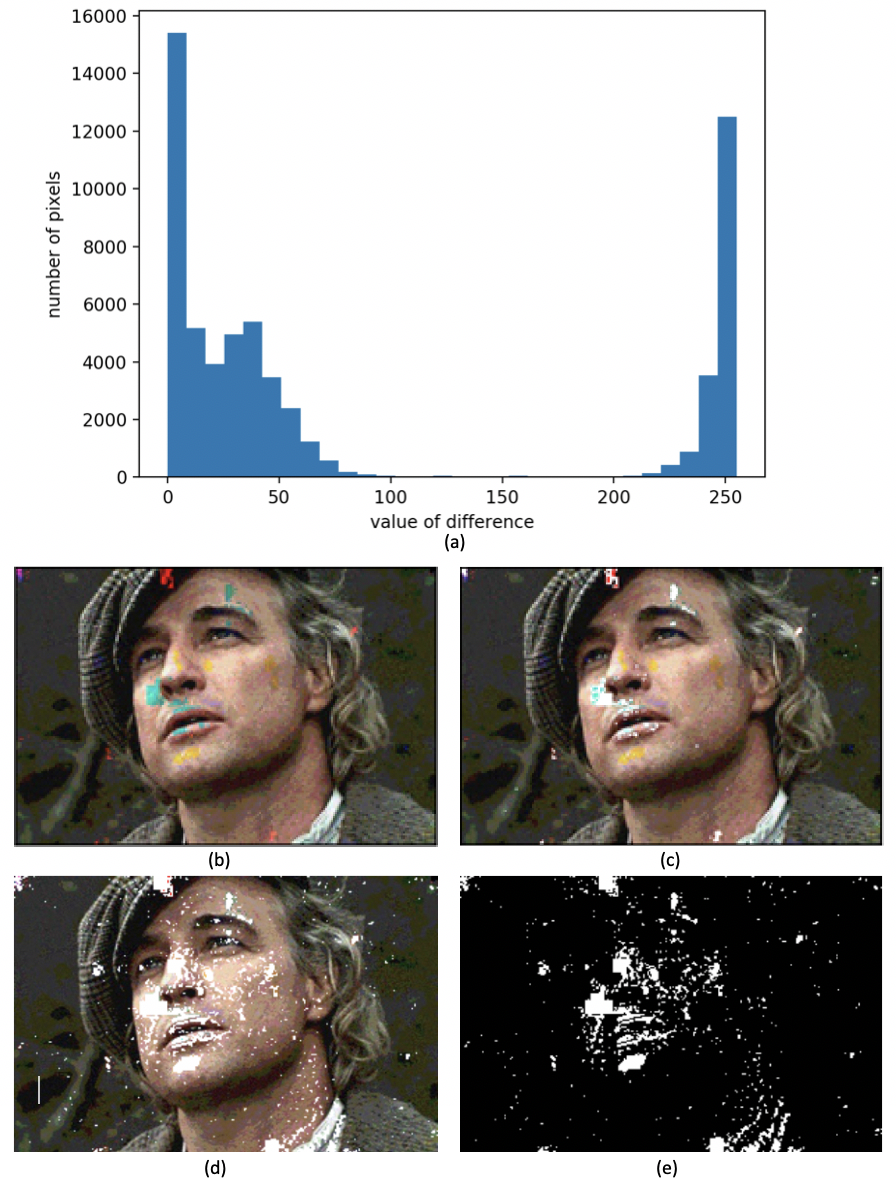}}
%\centerline{(a) Histogram of the values in the matrix $\textbf{R}-\textbf{G}$.}\medskip
% \centerline{\includegraphics[width=7.0cm]{hist_diff_2.png}}
% %\centerline{(b) Discoloration detection using differences between channels.}\medskip
% \centerline{\includegraphics[width=6.5cm]{mask.png}}
\vspace{-0.3cm}
\caption{(a) Histogram of the values in the matrix $\textbf{R}-\textbf{G}$. (b) Corrupted image. (c) Discolored regions in the red channel that have been whitened out. (d) An erroneous reconstruction with masking. (e) The Image of the mask.}

%Results of the discoloration detection algorithm.}
%\vspace{-0.5cm}
%\caption{Discoloration detection.}
\label{fig:hist_diff}
\end{figure}

\textbf{Automatic discoloration detection.} To the best of our knowledge, detecting arbitrarily shaped discolorations is a difficult problem in computer vision that has not been successfully addressed for classical image processing systems. This is due to the fact that discolored pixels usually have simultaneous distortions in all three color channels of possibly different degrees.
%So considering RGB channels jointly can only make the detection more challenging. Although discoloration is one of the most common seen defects in corrupted images, little work has considered this problem before, and no effective algorithm has been proposed to solve it so far.
However, detecting discolorations in DNA-encoded images is possible since with high probability, only one of the three color channels will be corrupted due to independent encoding of the RGB components. Figure~\ref{fig:rgb_sep} illustrates this fact, as erroneous pixels in different channels do not overlap. Within the correct color channels, pixels have neighbors of similar level, while within the erroneous channel, pixels have values that differ significantly from those of their neighbors. Figures~\ref{fig:hist_diff} (a)(b)(c) illustrates that pixels with the smallest $t=15$ frequencies in the difference vectors indeed correspond to almost all erroneous regions in the red channel.
%Our first known scheme for efficient automatic detection of discolorations in images is described in Algorithm~\ref{alg:detect}.
The results of our detection scheme are depicted in Figure~\ref{fig:hist_diff}(d)(e), for $t=18$. Note that the whitened out regions are treated as missing data, and filled in using inpainting techniques.

\begin{figure}[htb]
\centerline{\includegraphics[width=6.8cm]{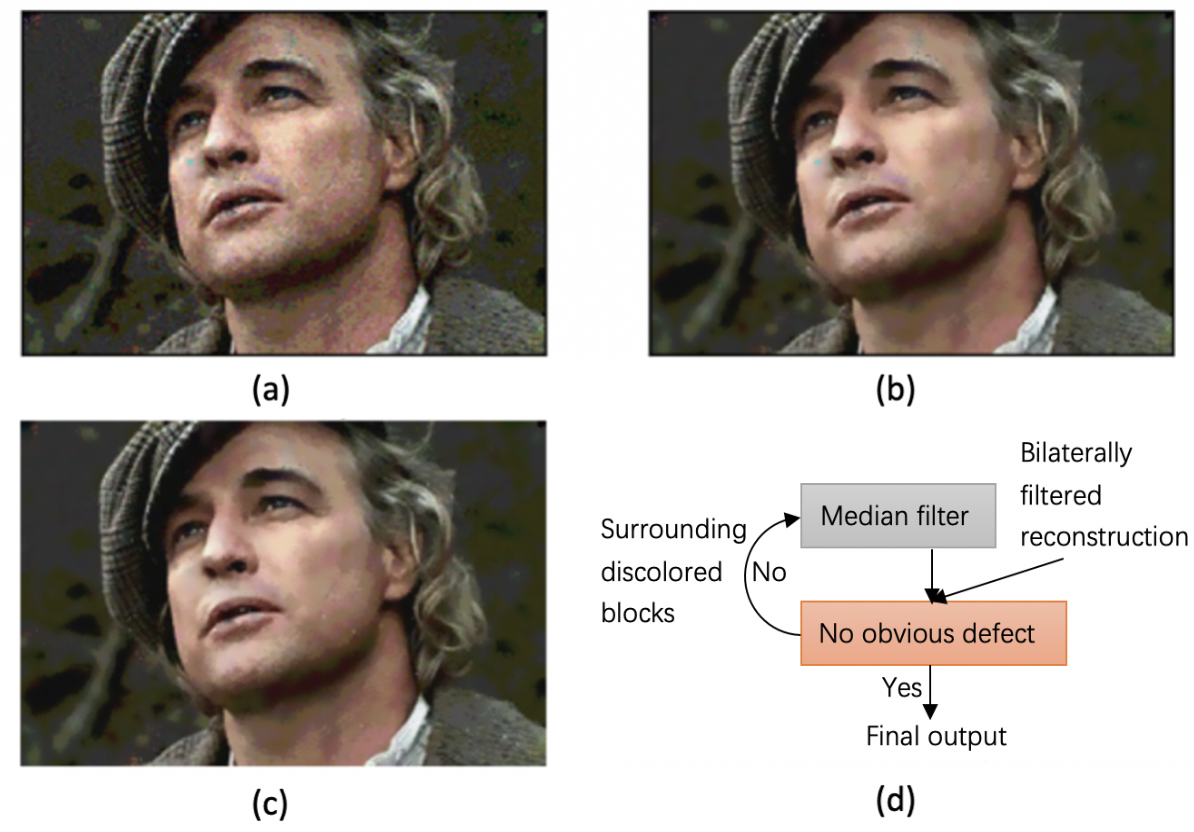}}
%\centerline{\includegraphics[width=12cm]{filter_2.png}}
%
\vspace{-0.5cm}
\caption{(a) Inpainting before smoothing. (b) Inpainting after smoothing. (c) Refined output of the inpainting procedure. (d) Description of the refining procedure.}
\label{fig:smooth}
\end{figure}

\textbf{Image inpainting} is a method for filling out missing regions in an image. There exist several methods for image inpainting currently in use, including diffusion-based, patch-based~\cite{bertalmio2000image} and deep learning approaches~\cite{Nazeri_2019_ICCV,yeh2017semantic,yu2019free}. The former two methods use local or non-local information only within the target image itself which leads to poor performance when trying to recover complex details in large images. On the other hand, deep-learning methods such as EdgeConnect \cite{Nazeri_2019_ICCV} combine edges in the missing regions with color and texture information from the remainder of the image to fill in the missing pixels. Since the encoded movie posters have obvious edge structures, we inpainted the images using EdgeConnect with the result shown in Figure~\ref{fig:smooth}(a). %As we can see, discoloration blocks are almost removed except for some small spots after inpainting.
% \begin{figure}[htb]
% \begin{minipage}[b]{.49\linewidth}
%   \centering
%   \centerline{\includegraphics[width=4.1cm]{4_recon.png}}
% \end{minipage}
% \hfill
% \begin{minipage}[b]{.49\linewidth}
%   \centering
%   \centerline{\includegraphics[width=2.5cm]{5_recon.png}}
% \end{minipage}
% % \centerline{\includegraphics[width=6.5cm]{4_recon.png}}
% % \centerline{\includegraphics[width=3.5cm]{5_recon.png}}
% %
% \caption{Reconstructed images using edge-connect.}
% \label{fig:edgeconnect}
% %
% \end{figure}

\textbf{Smoothing.} Although the problem of discoloration may be addressed through inpainting, the reconstructed images still suffer from mismatched inpaints and blocking effect caused by quantization. To further improve the image quality we perform smoothing through bilateral filtering~\cite{tomasi1998bilateral} that tends to  preserve the edges structures. The smoothing equations read as:
\begin{align}
\hat{I}[i,j] &= \frac{\sum_{[k,l]\in\Omega}I[k,l]w(i,j,k,l)}{\sum_{[k,l]\in\Omega}w(i,j,k,l)}, \notag \\
w(i, j, k, l) &=\exp \left(-\frac{(i-k)^{2}+(j-l)^{2}}{2 \sigma_{d}^{2}}-\frac{\|I[i, j]-I[k, l]\|^{2}}{2 \sigma_{r}^{2}}\right), \notag
\end{align}
where $I$ denotes the original image and $\hat{I}$ the filtered image, $\Omega$ is some predefined window centered at the coordinates $[i,j]$, and $\sigma_r$ and $\sigma_d$ are parameters that control the smoothing differences for intensities and coordinates, respectively. The filter performs Gaussian blurring on background regions but respects edge boundaries in the image. The result of smoothing with $\sigma_d^2=\sigma_r^2=45$ and $\Omega$ of the form of a $9\times 9$ square is shown in Figure~\ref{fig:smooth}(b), and no obvious discolorations are detectable. Furthermore, in order to address other possible impairments, we also used the positions of error blocks obtained from the discoloration detection platfrom to perform adaptive median smoothing around erroneous regions. The output of this iterative process is illustrated in Figure~\ref{fig:smooth}(c)(d).

%\textcolor{red}{Can you change the text in the refining procedure: Around discoloration blocks -> Surrounding discolored blocks?, Bilateral filtered reconstruction -> Bilaterally filtered reconstruction}
% \begin{figure}[htb]

% \centerline{\includegraphics[width=8.5cm]{bi_smooth.PNG}}

% %\centerline{\includegraphics[width=12cm]{filter_2.png}}
% %
% \caption{Example of smoothing.}
% \label{fig:smooth}
% %
% \end{figure}

% \begin{figure}[htb]

% \centerline{\includegraphics[width=8.5cm]{itr.PNG}}

% % \begin{minipage}[b]{.49\linewidth}
% %   \centering
% %   \centerline{\includegraphics[width=4.5cm]{iter_smooth_diag.PNG}}
% % \end{minipage}
% % \hfill
% % \begin{minipage}[b]{.49\linewidth}
% %   \centering
% %   \centerline{\includegraphics[width=4.0cm]{4_iter_smooth.png}}
% % \end{minipage}

% % \centerline{\includegraphics[width=6.5cm]{4_recon.png}}

% % \centerline{\includegraphics[width=3.5cm]{5_recon.png}}
% %

% \caption{Iterative refinement.}
% \label{fig:iter_smooth}
% %
% \end{figure}
%\section{Conclusion}
% \section{Experimental Results}
% \label{sec:exp}
% Need more image metric like PSNR and image similarity to this section.
%\vfill
%\pagebreak

\vfill\pagebreak

\bibliographystyle{IEEEbib}
\bibliography{refs}

\end{document}